%% file: paper.tex
\documentstyle[preprint,aps,prb]{revtex}
\tightenlines

\renewcommand{\appendix}{%
 \setcounter{section}{0}%
 \setcounter{equation}{0}%
 \renewcommand{\thesection}{{APPENDIX} \Alph{section}}
 \renewcommand{\theequation}{\Alph{section}.\arabic{equation}}%
}

\begin{document}
\preprint{LA-UR-98-4805}
\input zabst.tex
\input epsf.tex
\input sect1.tex

\input zref.tex
\input zfigs.tex

\input zfigps.tex
\end{document}

%% file: zabst.tex
\title{\bf Mesoscopic theory of the viscoelasticity of polymers}
\author{Shirish M. Chitanvis}
\address{
Theoretical Division, 
Los Alamos National Laboratory\\
Los Alamos, New Mexico\ \ 87545\\}

\date{\today}
\maketitle
\begin{abstract}
We have advanced our previous static theory of polymer entanglement
involving an extended Cahn-Hilliard functional, to include
time-dependent dynamics.
We go beyond the Gaussian approximation, to the one-loop level, to
compute the frequency dependent storage and loss moduli of the system.
The three parameters in our theory are obtained by fitting to available 
experimental data on polystyrene melts of various chain
lengths.
This provides a physical representation of the parameters in terms of the
chain length of the system.
We discuss the importance of the various terms in our energy
functional with respect to their contribution to the viscoelastic
response of the polymeric system.
\end{abstract}


%% file: sect1.tex

In a previous paper, we developed a static field theory of polymer
entanglement\cite{shirish2}, in which we introduced a non-local
attractive term, in addition to the usual excluded volume term, that
models  resistance to the motion of polymers due to entanglement.
Starting with this energy functional, we were able to use
Renormalization Group 
techniques to describe the onset of entanglement as the average molecular
weight is increased to a critical value.
The onset of entanglement may be described as a cross-over phenomenon, 
characterized by the effective diffusion constant going to zero as the 
transition point is approached, thereby implying critical slowing
down. 
We pointed out the existence of experimental evidence to support
the theory.

There have been several numerical approaches developed to understand the
viscoelastic response of polymers\cite{termonia,bicerano,holtzl}.
It is of interest to see whether an alternative theory of viscoelasticity
of polymers using continuum
concepts can be developed.
Our previous theory, being static in nature, clearly needs to be
extended if one is to
study the time-dependent response of polymeric systems.
The chief purpose of this paper is to lay down the foundations of a
time-dependent field theory of entangled polymers, primarily through a 
comparison with experimental results on the linear viscoelastic
response of polymer melts.
In future work, we shall probe the time-dependent approach to entanglement of
polymeric systems, as the molecular weight is increased to a critical
value.
In this, we have in mind an analogy with dynamic critical phenomena,
where one of the quantities of interest is the frequency dependent
diffusion constant, and the manner in which it scales to zero as the
transition is approached. We will also compare the results of our theory
briefly to the results of the standard reptation theory
approach\cite{doied}.

The time-dependent internal energy functional $U$ which extends our
previous static theory\cite{shirish2} can be written down in a
straightforward manner in terms of an energy density $u$:

\begin{eqnarray}
&&U = \int d^3s \int dt~  u({\bf s},t) \nonumber\\
&&\beta u = 
 c({\bf s},t) {\partial c({\bf s},t) \over \partial t } \nonumber\\
&&+ \left({\alpha\over \sqrt 2}\right) 
      {{\partial {c}({\bf s},t)}\over{\partial s_i}}
                          {{\partial {c}({\bf s},t)}\over
{\partial s_i}} \nonumber\\
  &&+\left({\alpha^2\over 2}\right) 
  ~{{c}}({\bf s},t) {{c}}({\bf s},t) \nonumber\\
     &&- \left({\alpha^4 \over {2 \pi}}\right) \int {\rm d}^3 s'
                     {{c}}({\bf s},t) {exp(-\delta
                                        \vert {\bf s} - {\bf s}' \vert)\over
          {\vert {\bf s} - {\bf s}' \vert} }~   {{c}}({\bf s}',t)
        \nonumber\\
\beta && = {1 \over k T}
\label{visco}
\end{eqnarray}

where $c$ is the number concentration of the polymer strands, and
$\delta^2 = \sqrt 2 \alpha$, k is Boltzmann's constant and T is the 
temperature.
This model is an extension of the standard Cahn-Hilliard approach\cite{cahn}.
The first two terms in Eqn.(\ref{visco}), taken by themselves, are
easily seen to lead to the time-dependent linear diffusion equation,
with $\alpha$ plays a role analogous to that of a diffusion constant.
The third term in the equation represents the standard excluded volume 
interaction.
We will discuss $\alpha$ in greater detail shortly.
The final non-local attractive term represents the fact that when
polymers become entangled, there will be in general a resistance to
their movement.
Similar approaches can be found in the literature, where authors have
taken activation energies to represent entanglement\cite{bicerano}.
The form of our interaction term is novel, however, and we have
discussed in the previous paper\cite{shirish2} the rationale for
choosing the parameter $\alpha$ in the model is as given in that
paper, since it leads to a number preserving Euler-Lagrange equation.
It is worth noting that even standard treatments involving the
excluded volume term are generally restricted to the static case,
whereas we have considered here an extension to the frequency-dependent 
case as well.

The time and space co-ordinates in the equation are dimensionless, the 
scales we have chosen being $\omega_c^{-1}$ and $\lambda$
respectively.
The dimensionless parameter $\alpha$ can be written as
$D/[\lambda^2 \omega_c]$, where $D$ is the diffusion constant which
scales as the inverse square of the molecular weight in the entangled state.
Since both $\lambda$ and $\omega_c$ could possess a molecular
weight dependence of their own, 
it follows that $\alpha$ can display a dependence on molecular
weight different than the scaling behavior of $D$.

Since we wish to study polymer melts undergoing shear experiments,
where they are in contact with an energy reservoir (at constant
temperature), the energy which is conserved is the Helmholtz free
energy $A = U - ST$\cite{callen}, $S$ being the entropy and $T$ the
temperature. 
The entropy is given approximately as\cite{raveche}:

\begin{eqnarray}
S &&= \int d^3s \int dt { s}({\bf s},t) \nonumber\\
{ s}({\bf s},t) &&\approx c({\bf s},t) \ln[c({\bf s},t)] \nonumber\\
(1+c') \ln (1+c') &&\approx 
        c' 
     + {c'^2\over 2 } - {c'^3 \over 6 } +{c'^4 \over 12}
\label{ent}
\end{eqnarray}

The last of these equations indicates an expansion around
$\lambda^{-3}$, which is identical to one in the units we have chosen.
The linear terms will be ignored following convention, as they can be
absorbed into the chemical potential $\mu$, required for number
conservation.
In the mean field approximation, $\mu \equiv 0$.
In what follows, we shall drop the primes on the number concentration
variable. 

The goal of this communication is to compute the linear viscoelastic
response of a polymeric system.
This can be done following closely the analysis in our previous
paper\cite{shirish2}, to obtain an expression for the frequency
dependent stress $\sigma(\omega)$:

\begin{eqnarray}
\sigma(\omega) &&= -i \omega~C {\cal S}(k=0,\omega) \epsilon(\omega)
\nonumber\\ 
C &&= \left( {k T \over  \omega_c \lambda^3} \right)
\label{stress}
\end{eqnarray}

where $S(k,\omega)$ is the two-point correlation function for the
system.
Note that since we chose to take temporal Fourier transforms with
respect to $\exp(-i \omega t)$, our sign convention in the first of
Eqns.(\ref{stress}) is opposite that in standard literature\cite{doied}.
In general $S(k,\omega)$ is given by:

\begin{eqnarray}
{\cal S}(k,\omega) &&= \left({\cal
    S}_0^{-1}(k,\omega)-\Sigma(k,\omega)\right)^{-1} 
\nonumber\\
{\cal S}_0(k,\omega) &&= \left(-i \omega +s_0^{-1}(k) \right)^{-1}
\nonumber\\
s_0(k) &&= \left(1 + \sqrt 2 \alpha k^2 + 2 \alpha^2 k^2/
(1+\sqrt 2 k^2/\alpha)\right)^{-1} 
\approx \left(1 + a k^2 \right)^{-1} \nonumber\\
a &&= 2 \sqrt 2 \alpha
\label{vis-form}
\end{eqnarray}

where, as usual $\Sigma$ denotes the self-energy.
From Eqn.(\ref{vis-form}), we see that $a$ plays the role of a
diffusion constant.
But bearing in mind the discussion below Eqn.\ref{visco}, we expect
its dependence on molecular weight to be different than the
conventional 
diffusion constant $D$, due to the manner in which we have scaled our
variables. 

From Eqns.(\ref{stress}) and (\ref{vis-form}), we see that we need to
evaluate the correlation function in the long wavelength limit.
We can perform this calculation using a perturbation expansion with
respect to the 
nonlinear terms, using standard methods from field theory.
These are elementary extensions of the methods detailed in Ref. 1.
The vertices we obtain from Eqn.(\ref{ent}) 
are depicted in Fig.1\cite{ramond,binney}.
The diagrams which we consider are depicted in Figures 2 and 3.
It is easy to show  that the tadpole and bubble diagrams
vanish identically and the only surviving lowest order diagram is the
setting-sun diagram, whose contribution can be shown analytically to
be:

\begin{eqnarray}
\Sigma_{2b} (\omega,k=0,\omega_m) &&= {1\over 4}~\
\int {d\omega'\over2 \pi}~\int {d^3 k \over (2 \pi)^{3}}~\
{\cal S}_0(k,\omega')~{\cal S}_0(k,\omega-\omega') \nonumber\\ 
 =&& \
{ \sqrt{i \omega - 1} \over {32 \pi^2 a^{3/2}}}~ \ 
\left[
\ln \left[ {(\omega/2 +\omega_m)\over (\omega/2-\omega_m)} ~{(3 \omega/2 - \omega_m)\ \over (3 \omega/2+ \omega_m)} \right]\right]
\label{sigma}
\end{eqnarray}

Here, $\omega$ is the frequency at which the system is being sheared.
In order to obtain the final expression for the self-energy in the
long wavelength limit,
we first performed the frequency integral.
This integral is logarithmically divergent, rendered finite
through the use of a high-frequency cut-off $\omega_m$, using the
conventional procedure from field theory.
$\omega_m$ may be interpreted in the usual field-theoretic sense as an 
upper frequency limit below which the continuum theory is not valid.
The dependence of the final answer on the cutoff
$\omega_m$ is logarithmic.
Next, to perform the k-integration,
we used the method of
contour integration, taking care to distort the contour to avoid the
branch cut implied by the logarithmic behavior of the integrand in
k-space.
Once this is done, 
it is straightforward to use the method of residues 
to get the expression given above.
It is easy to verify that the integrand in k-space is indeed suitably
convergent, and no additional cutoffs are necessary.

With Eqns.(3)-(5) in hand, we fitted experimental data on polystyrene
melts characterized by a very low polydispersity, and a fairly wide range of
molecular weights, ranging from just under $10^4$ to about
$580,000$\cite{onogi}.  
The results are encapsulated in Fig. 4.
Three sets of data are shown in the figure, chosen to represent low,
medium and high molecular weight samples.
We have performed the fitting procedure for all the sets of data
provided by Onogi et al\cite{onogi}.
We have chosen to display three of the fits as being representative of 
the procedure.
We see that the fits, while good, show signs of deteriorating slightly 
as the molecular weight decreases.
The plateaus indicate the rubbery phase of the system.
Our previous paper applies in this region\cite{shirish2}.
For higher frequencies (short times), the stress is much higher, which 
could be interpreted in terms of inertia as the system starts to be
strained.  For frequencies below the plateaus (long-time behavior),
one might say that the polymers eventually begin to disentangle,
causing the stress to start decreasing precipitously.

We began with four parameters in the theory, viz., $\lambda$,
$\alpha$, $\omega_m$ and $\omega_c$.
In performing the fitting, due to the manner in which the parameters
appear in the expression for the complex modulus,
it was more convenient to use the reduced set of three parameters
$C, \omega_m, a$, where $C$ and $a$ have both been defined earlier.
While it is true that the parameters in our theory had to be chosen to
get the best fit with data, we find it 
impressive that the function given in
Eqn.(\ref{sigma}) is such that it provides the correct {\it form} for
the storage modulus.
In this manner, our theory has captured the essential aspects of the
linear viscoelasticity of polymer melts.
From the values obtained for the three parameters, we were able to
perform a least-squares fit, yielding the following representations as 
functions of the molecular weight $M_n$.

\begin{eqnarray}
C &&\approx 3.1 \times 10^{-13}~ M_n^{3.7} ~ (dyne-s-cm^{-2}) 
\nonumber\\  
\omega_m &&\approx b_0~M_n^{b_1}~(s^{-1});~b_0 = 0.11;~b_1 = -1.53 
\nonumber\\  
a &&\approx d_0~M_n^{d_1};~d_0 = 0.21 \times
10^{-13};~d_1 = 1.35 
\label{param}
\end{eqnarray}

Note the scaling forms for these representations.
First of all, as will be shown shortly, $C$ can be
identified with the static viscosity, and we found the scaling
exponent $3.7$ to be reasonably close to the value of $3.4$ given in
the literature.
To obtain this representation, it was useful to plot the fitted values 
of $C$ against the molecular weight $M_N$ on a log-log plot.
The cut-off frequency $\omega_m$ scales inversely as approximately the 
$3/2$ power of the molecular weight.
Finally, as advertised earlier, the dependence of $a$ on $M_n$ is
different than that of the true diffusion constant $D$.
The reason for this is that $a$ is proportional to $D \lambda^{-2}
\omega_c^{-1}$, so that the product
$\lambda^{2}\omega_c$, which has the units of a diffusion
constant, goes approximately as $ \sim M_n^{-10/3}$.
It is further easy to show that $\lambda \sim M_n^{-2/5}$, and 
$\omega_c \sim M_n^{5/2}$.

As one might perhaps expect, the entropy terms in our energy
functional have a dominant effect on determining the linear
viscoelastic
behavior of polystyrene melts.
The nonlocal attractive term, which models the resistance to the
motion of entangled polymers has a less pronounced effect on
linear viscoelasticity.
This is consistent with our earlier calculations in the static
regime\cite{shirish2}, where we found that the nonlocal attractive
term has a more profound effect on determining the renormalized
diffusion constant than the elastic moduli.
We will tackle the frequecny dependence of the renormalized diffusion
constant in future work.

From Fig. 4 we see that the fits to data using our theory are quite
good, as are those using the standard reptation theory\cite{doied}.
The advantange of our theory is that while the reptation theory is
restricted to the regime of highly entangled systems, and uses
mean-field 
concepts, we can compute the effect of fluctuations
using Feynman diagrams.
With these parameteric representations, we were also able to compute
the loss moduli for the samples. 
The curves gave a reasonable but only an average fit for the various
samples,
In other words, the loss moduli obtained through our procedure was not 
very sensitive to the parameters we obtained.
Nevertheless, we note that this is an improvement over the standard
reptation model approaches, which give a null loss
modulus\cite{doied}.
We also went a bit further, and attempted to use the above
representations to calculate the storage moduli of another polymer
melt, and found reasonable agreement, suggesting that our approach has 
a universal flavor to it.
On the other hand, the results for solid polymers were quite abysmal,
indicating that our fitting procedure must be redone for non-melts.
We have been unable to locate data for solid polymers or other polymer 
melts having the same wide scope as the results
of Onogi et al\cite{onogi} for polystyrene melts.

It is easy to show that in the zero frequency limit, the viscosity is
given by:

\begin{equation}
\eta_0 = Lim_{\omega \to 0}~ \omega^{-1} Im \left[-i \omega C {\cal
    S}(k=0,\omega) \right]\equiv C
\label{lim1}
\end{equation}

One similarly obtains an expression for the shear modulus in the zero
frequency limit by considering the storage modulus.
For sufficiently large values of the frequency, the experimental data
show an approximately linear behavior, as does the present theory.

As mentioned at the beginning, the purpose of this note is to lay down 
the foundations of a time-dependent field-theoretic approach to the
viscoelastic response of polymeric systems.
Our eventual goal is to understand the dynamic approach to
entanglement, analogous to treatments of dynamic critical
phenomena\cite{hoh}.

I would like to acknowledge a useful comment by Sanat Kumar
concerning the calculations.
This research is supported by the Department of Energy contract
W-7405-ENG-36, under the aegis of the Los Alamos National Laboratory
LDRD polymer aging CD program.


%% file: zref.tex

%% file: zfigs.tex
\begin{figure}
\caption{(a) is a pictorial representation of the cubic term in $A$.
Each leg corresponds to a factor of $c$, the field.
The intersection of the three legs symbolizes a factor of $\gamma=1/6$, 
the coupling constant.
(b) is a pictorial representation of the quartic term in $A$.
A factor of $-1/12$ is to be inserted at the intersection.}
\label{fig1}
\end{figure}
\begin{figure}
\caption{ (a) represents the {\it tadpole} diagram which is crucial in our
calculations.
(b) represents the {\it setting sun} diagram.
Both (a) and (b) are second order contributions to the correlation
function coming from the cubic interaction term, the first order
corrections being null.}
\label{fig2}
\end{figure}
\begin{figure}
\caption{ This figure represents 1-loop (bubble) contribution from the quartic 
interaction term in $A$.}
\label{fig3}
\end{figure}
\begin{figure}
\caption{ This plot shows a comparison of our theory (solid line) with 
  the experimental data by Onogi et al on polystyrene melts.  The
  molecular weights are placed alongside the different sets.
The comparison is fairly good, and appears to deteriorate slightly as
the molecular weight decreases, indicating that the theory works
better for high molecular weight melts.
The plateaux indicate the rubbery regime for each sample
}
\label{fig4}
\end{figure}
\newpage


%% file: zfigps.tex
\begin{figure}

\centerline{Chitanvis,\ \ Fig.\ \ref{fig1}}
\epsfxsize=7cm
\centerline{\epsffile{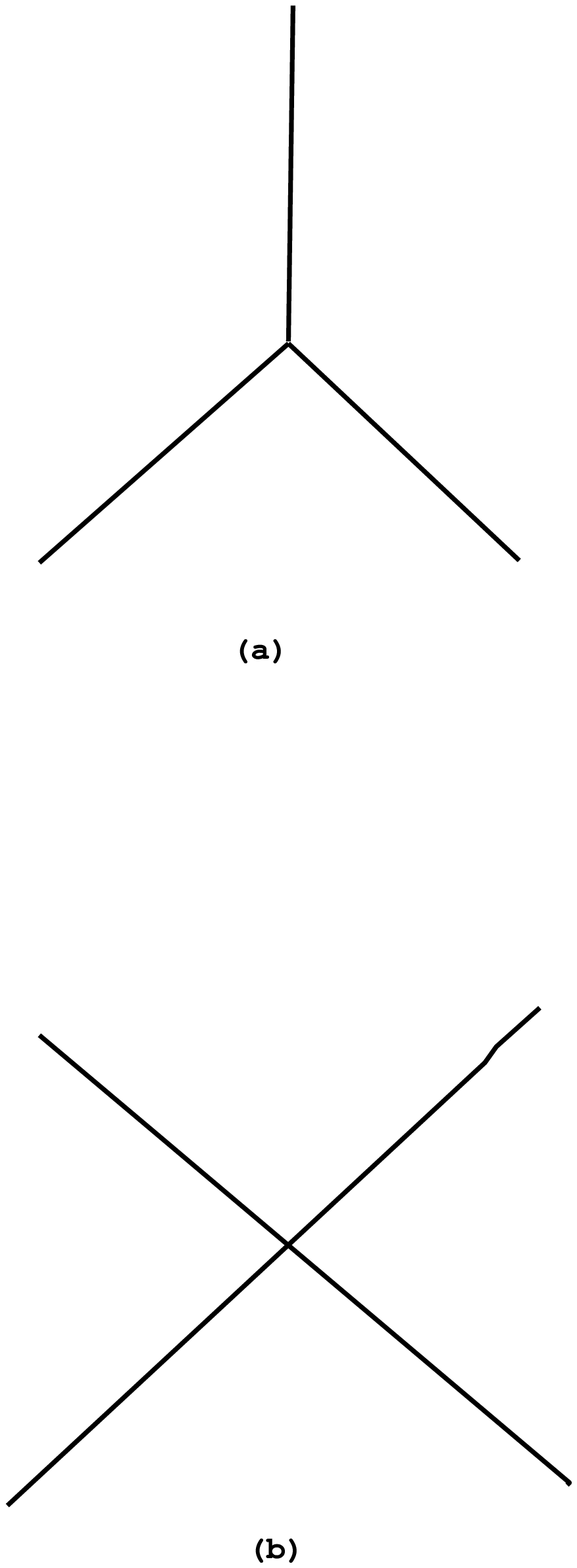}}
\eject
\end{figure}

\begin{figure}
\centerline{Chitanvis,\ \ Fig.\ \ref{fig2}}
\epsfxsize=10cm
\centerline{\epsffile{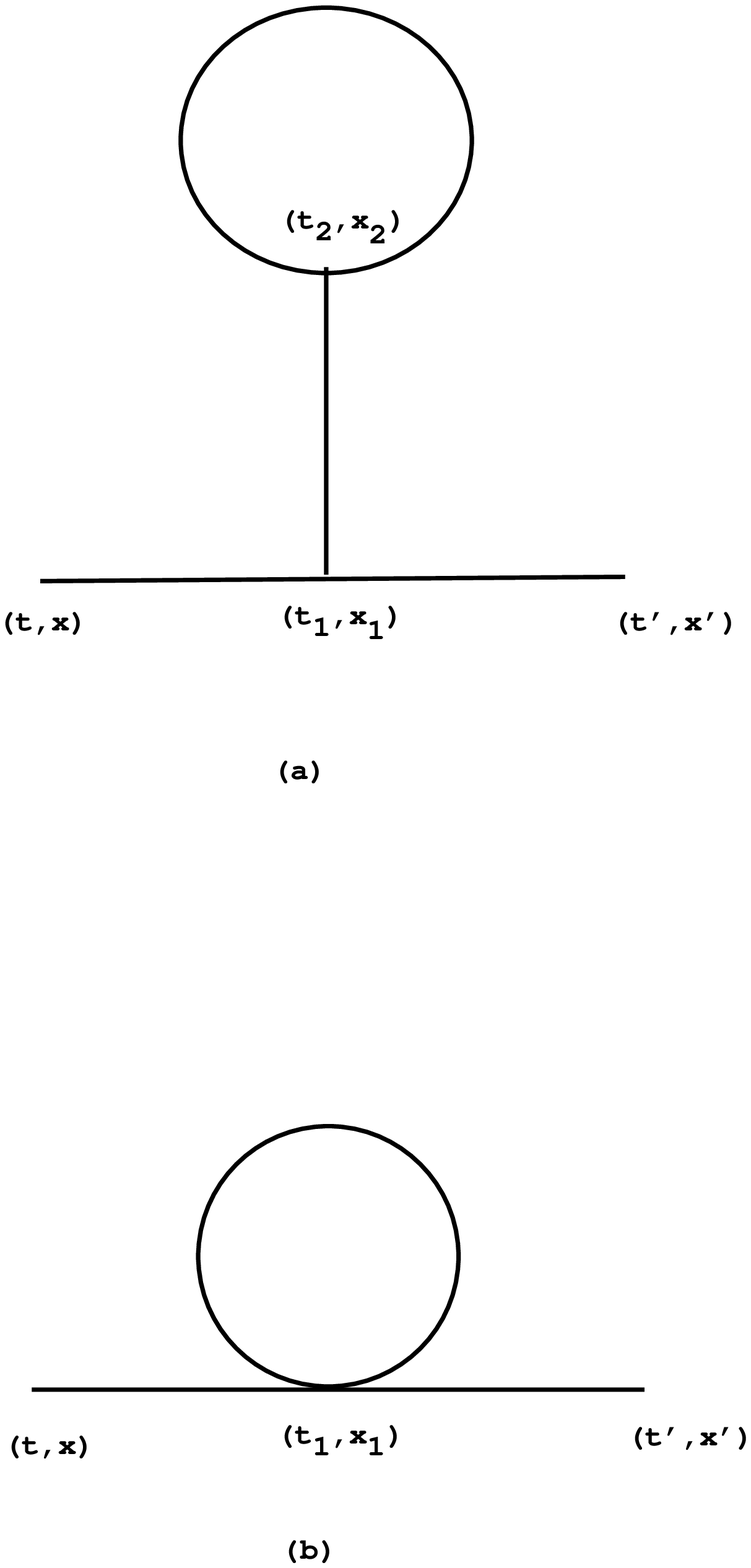}}
\vfill\eject
\end{figure}

\begin{figure}
\centerline{Chitanvis,\ \ Fig.\ \ref{fig3}}
\vspace{5cm}
\epsfxsize=15cm
\centerline{\epsffile{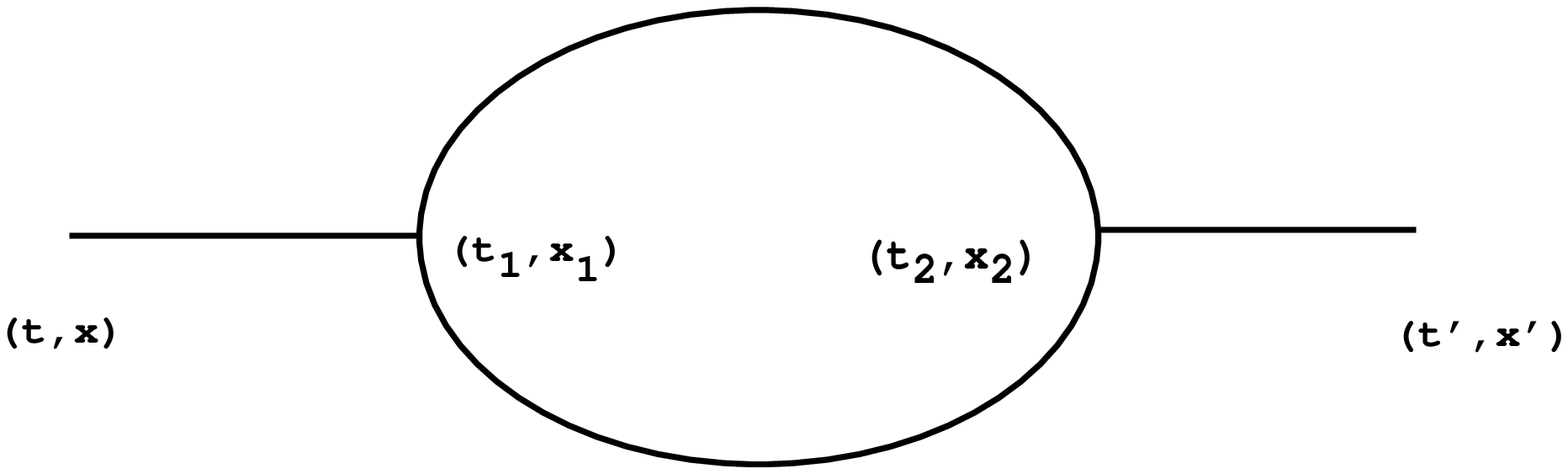}}
\vfill\eject
\end{figure}

\begin{figure}
\centerline{Chitanvis,\ \ Fig.\ \ref{fig4}}
\vspace{5cm}
\epsfxsize=20cm
\centerline{\epsffile{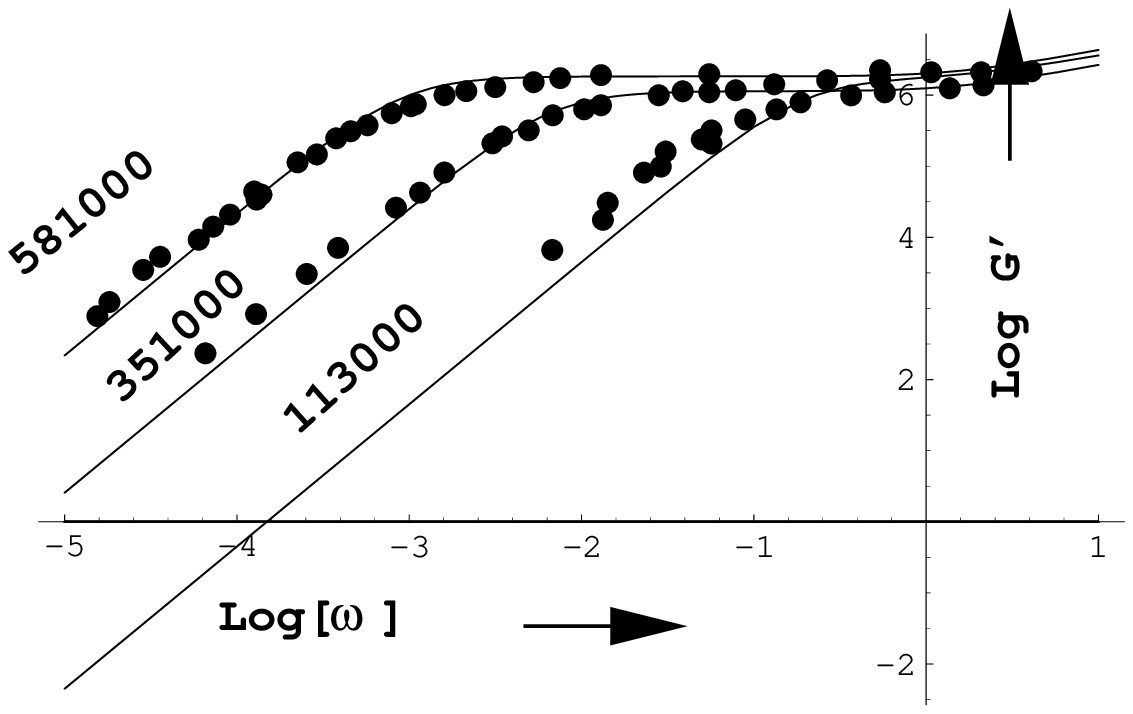}}
\end{figure}
